# A 28/37/39GHz Multiband Linear Doherty Power Amplifier in Silicon for 5G Applications

Song Hu, *Member, IEEE*, Fei Wang, *Student Member, IEEE*, and Hua Wang, *Senior Member, IEEE*

*Abstract*—This paper presents the first multiband mm-wave linear Doherty PA in silicon for broadband 5G applications. We introduce a new transformer-based on-chip Doherty power combiner, which can reduce the impedance transformation ratio in power back-off (PBO) and thus improve the bandwidth and power-combining efficiency. We also devise a "driver-PA co-design" method, which creates power-dependent uneven feeding in the Doherty PA and enhances the Doherty operation without any hardware overhead or bandwidth compromise. For the proof of concept, we implement a 28/37/39-GHz PA fully integrated in a standard 130-nm SiGe BiCMOS process, which occupies 1.8mm². The PA achieves a 52% −3-dB small-signal $S_{21}$ bandwidth and a 40% −1-dB large-signal saturated output power ($P_{sat}$) bandwidth. At 28/37/39GHz, the PA achieves +16.8/+17.1/+17-dBm $P_{sat}$, +15.2/+15.5/+15.4-dBm $P_{1dB}$, and superior 1.72/1.92/1.62 times efficiency enhancement over class-B operation at 5.9/6/6.7-dB PBO. Moreover, the PA demonstrates multi-Gb/s data rates with excellent efficiency and linearity for 64QAM in all the three 5G bands. This PA advances the state of the art for Doherty, wideband, and 5G silicon PAs in mm-wave bands. It supports drop-in upgrade for current PAs in existing mm-wave systems and opens doors to compact system solutions for future multiband 5G massive MIMO and phased-array platforms.

*Index Terms*—5G, broadband, integrated circuits, Doherty, efficiency, linearity, massive MIMO, multiband, phased array, power amplifier, power back-off, reconfiguration, silicon, SiGe BiCMOS, transformer.

## I. INTRODUCTION

MULTIPLE use cases in future fifth-generation (5G) wireless systems, such as enhanced mobile broadband (eMBB), target multi-Gb/s user-experienced data rates [1]. Such ultra-high throughput will not only augment existing wireless data communication, but also enable numerous emerging applications such as augmented reality (AR), virtual reality (VR), and mixed reality (MR) [2], [3].

To achieve multi-Gb/s data rates, the 5G wireless systems are evolving towards mm-wave [4], [5]. The Federal Communications Commission (FCC) has opened multiple mm-wave frequency bands for 5G development in the United States, including spectra around 28, 37, and 39GHz [6]. Different regions in the world are interested in various mm-wave frequency bands for 5G (Fig. 1) [6], [7]. Moreover, the mm-wave 5G systems will extensively leverage massive multiple-input multiple-output (MIMO) and phased-array architectures to improve the link performance by strengthening desired signal and enhancing interference rejection. As a result, multiband mm-wave 5G systems are highly desired to facilitate future international/cross-network roaming and MIMO diversity. Together with existing wideband mm-wave antenna arrays [8]-[10], multiband mm-wave 5G circuits will greatly shrink the form factor of future massive MIMO and phased-array 5G systems. On top of the evolution towards mm-wave, the future 5G systems will leverage spectrum-efficient modulations such as high-order quadrature amplitude modulations (QAMs) to achieve high data rates for given spectrum resource [1]. These 5G waveforms have large peak-to-average power ratios (PAPRs) and lead to circuit design challenges [11]. Mobile devices utilizing mm-wave 5G techniques, such as 5G handsets and VR headsets, should be energy efficient in power back-off (PBO) to extend the battery life and ease the thermal management. Also due to the high PAPR, these mm-wave 5G systems need to be highly linear in a wide power range for both their amplitude and phase responses to ensure the quality of service and thus user experience.

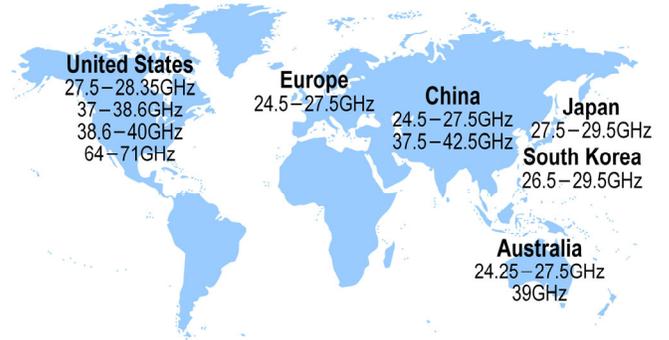

Fig. 1. Global spectrum allocations for mm-wave 5G.

Power amplifier (PA) is often the most power-hungry block in a wireless transceiver and need to handle large signals. Therefore, PA often governs the energy efficiency and linearity of a wireless transmitter [12], [13]. Significant progresses in the silicon-based mm-wave PA have been made in the past decade [14]-[43]. Wideband mm-wave PAs using distributed topologies [17] or high-order passive networks [30] have been demonstrated in literature. However, they often compromise efficiency due to inefficient active circuits or passive networks. Advanced distributed PA designs improve the efficiency with increased overhead in the power management unit [32]. In parallel, various PA PBO efficiency enhancement techniques have been demonstrated at mm-wave. However, many of them entail challenges for the large modulation bandwidth in broadband 5G applications. Mode-switching PAs [26], [36] entail challenges to achieve dynamic power-mode switching at the speed of 5G signals' envelopes. Outphasing PAs require the generation of high-speed outphasing signals, which demands significant baseband overhead [23]. Envelope tracking (ET) PAs require high-speed supply modulators with wide dynamic ranges, and state-of-the-art ET designs demonstrate modulation bandwidth only up to tens of MHz [44], [45]. Compared with others, Doherty PAs fundamentally can support the wide modulation bandwidth in broadband 5G applications [46]-[51]. However, Doherty PAs require complex power combiners, which could be lossy and limit the carrier bandwidth. In addition, Doherty PAs demand careful cooperation between at least two PA paths to ensure the efficiency and linearity. Although digital-intensive Doherty PAs in recent research address this issue at sub-6GHz [52]-[59], their direct extension to mm-wave broadband 5G could result in unaffordable baseband overhead. Consequently, existing mm-wave Doherty PAs in silicon exhibit limited PBO efficiency enhancement.

Therefore, wideband mm-wave PA in silicon with efficiency enhancement is an urgent-yet-unmet need for 5G massive MIMO and phased-array systems. We demonstrate the first mm-wave linear Doherty PA in silicon, which achieves substantial PBO efficiency enhancement and supports three mm-wave 5G bands with multi-Gb/s data rates [60]. Section II and III present the introduced transformer-based low-loss/broadband Doherty power combiner and power-dependent Doherty PA uneven-feeding scheme, respectively. They are two enabling techniques that address the challenges in conventional mm-wave Doherty PA designs. Sections IV shows the implementation details and test results.

## II. Transformer-Based Low-Loss and Broadband Doherty Power Combiner

Doherty PA is generally comprised of multiple PA active cells, a Doherty power combiner, and an input network (Fig. 2a). The input network balances the phases of PA paths to ensure efficient power combining at the PA output. The Doherty power combiner plays critical roles. It performs not only power sum, but also active load modulation that enhances the efficiency of PA active circuits. Its performance has large impacts on the PA efficiency and bandwidth and often dominates the chip area in fully integrated Doherty PAs. This section first reviews the conventional approaches of two-way Doherty power combiners and then introduces a new low-loss, broadband, and compact design.

In the following analyses, we assume that the main and auxiliary PAs share the same supply voltage, and the maximum RF current of the auxiliary PA is α times the maximum RF current of the main PA[1]. In a symmetric Doherty PA, α equals one, and the second efficiency peak ideally happens at 6-dB PBO. In asymmetric Doherty PAs, the second efficiency peak happens at <6-dB PBO or >6-dB PBO when the auxiliary PA is weaker (α<1) or stronger (α>1) than main PA. We derive the desired relationship between the RF currents of the main and auxiliary PAs as

$$i_{aux} = \begin{cases} (1+\alpha)i_{main} - \dfrac{2}{1+\alpha}, & \dfrac{2}{(1+\alpha)^2} \le i_{main} \le \dfrac{2}{1+\alpha} \\ 0, & 0 \le i_{main} < \dfrac{2}{(1+\alpha)^2} \end{cases}, \quad (1)$$

where $i_{main} \in [0, \ 2/(1+\alpha)]$ and $i_{aux} \in [0, \ 2\alpha/(1+\alpha)]$ are the normalized RF currents of the main and auxiliary PAs [61]. Fig 2b illustrates the symmetric case (α=1). By these definitions, the sum of maximum $i_{main}$ and $i_{aux}$ is independent to α, and the Doherty PA peak output power ($P_{out}$) remain constant when α varies. The PBO level in decibel is calculated as $20\log_{10}\{2/[(1+\alpha)i_{main}]\}$ [61], and the load-pull impedances of the main and auxiliary PAs at the Doherty PA peak $P_{out}$ are defined as $(1+\alpha)R_{opt}/2$ and $(1+\alpha)R_{opt}/(2\alpha)$, respectively.

### A. Conventional Doherty Power Combiners
#### 1) Parallel Doherty Power Combiners
Most conventional Doherty power combiners derive from a two-λ/4-line-based architecture in Fig. 3. These Doherty power combiners perform parallel power combining for current-mode PAs and achieve true Doherty load modulation.

In this architecture, the λ/4 line TL₂ down scales the PA load impedance for high PA $P_{out}$. It transforms $R_L$ to $R_{opt}/2$ at all PA $P_{out}$ levels, and its characteristic impedance is

$$Z_{02} = \sqrt{\dfrac{R_{opt}R_L}{2}}. \quad (2)$$

The other λ/4 line at the main PA output, TL₁ in Fig. 3, performs as an impedance inverter, and its characteristic impedance is derived as

$$Z_{01} = \frac{1+\alpha}{2} R_{opt} . \quad (3)$$

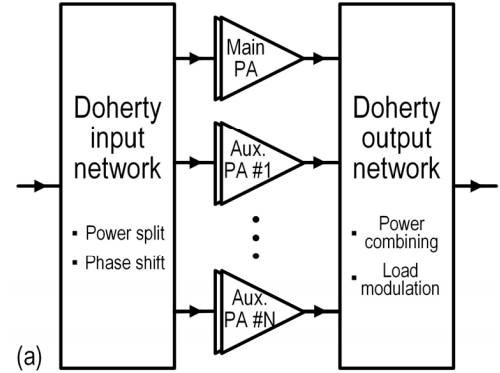

(a)

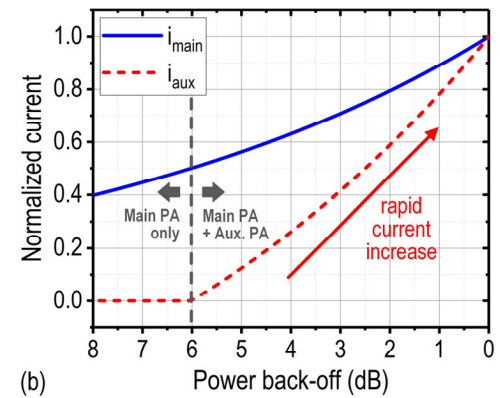

(b)

Fig. 2. (a) Conceptual diagram of a Doherty PA. (b) Ideal cooperation between the main and auxiliary paths in a symmetric two-way Doherty PA.

---

[1] If not otherwise specified, the RF voltage and current in our discussions are for the component at the fundamental frequency.

To implement these $\lambda/4$ lines in silicon, conventional integrated designs often employ transmission lines [27], $\pi$-network approximations using lumped elements [62]-[65], or transformer-based solutions [54], [55]. However, any implementations derived from this two-$\lambda/4$-line-based Doherty power combiner inherently suffer high impedance transformation ratios (ITRs) in PBO for the impedance inverter. When the auxiliary PA is on, i.e., $i_{main} \in [2/(1+\alpha)^2, 2/(1+\alpha)]$, the ITR of TL$_1$ in Fig. 3 is derived as

$$ITR_{Conv} = [\frac{1+\alpha}{(2+\alpha) - \frac{2}{1+\alpha}/i_{main}}]^2 . \quad (4)$$

Note that the term in the square bracket in (4) is always $\geq 1$. (4) implies two important mathematical insights. First, for any given $\alpha > 0$, ITR$_{Conv}$ monotonically increases when $i_{main}$ decreases. In other words, the ITR of TL$_1$ in Fig. 3 gradually increases during PBO, until the auxiliary PA is off (Fig. 4). For example, ITR$_{Conv}$ in the symmetric design is unity at the peak P$_{out}$ and is as high as four times at 6-dB PBO. Note that such high ITR in PBO is independent of R$_{opt}$ and thus the Doherty PA peak P$_{out}$. To show the second insight, we substitute the $i_{main}$ value at the second efficiency peak, $2/(1+\alpha)^2$, into (4), and get ITR$_{Conv} = (1+\alpha)^2$. This result tells that a stronger auxiliary PA always leads to a larger ITR in PBO (Fig. 4).

For a $\lambda/4$ line performing impedance transformation, a larger ITR results in a higher loss and narrower bandwidth [66]-[70]. Therefore, the two mathematical insights into (4) lead to two physical drawbacks of the two-$\lambda/4$-line-based Doherty power combiner and its derivatives. First, due to the increasing ITR in PBO, the passive efficiency degrades in PBO, and the bandwidth is also compromised. Second, the passive efficiency degradation and bandwidth compromise get worse when the auxiliary PA is becoming stronger. An asymmetric Doherty PA with a stronger auxiliary path may fit the PBO efficiency profile with the power probability density functions (PDFs) better for high-PAPR signals and thus enhances the average efficiency [47]. However, the second drawback of the two-$\lambda/4$-line-based Doherty power combiner compromises the asymmetric design techniques. Note that these deteriorations in passive efficiency and bandwidth are independent to the Doherty PA peak P$_{out}$.

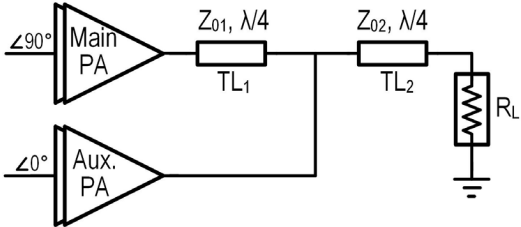

Fig. 3. Two-$\lambda/4$-line-based parallel Doherty power combiner.

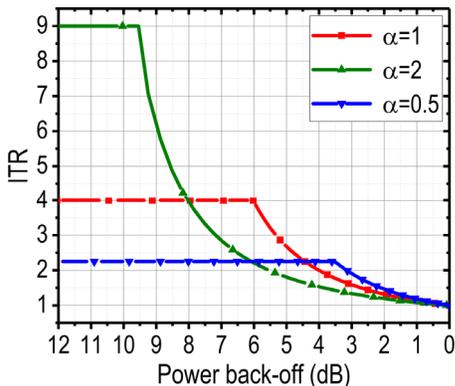

Fig. 4. ITR of the impedance inverter in the two-$\lambda/4$-line-based parallel Doherty power combiner.

### 2) Series Doherty Power Combiners

Researchers demonstrate Doherty PAs that use series power combiners summing the power from current-mode PAs and achieve true Doherty load modulation. However, in order to properly terminate the series-combining transformer at the auxiliary PA side when the auxiliary PA is off, additional matching networks [29], [65], [66] or switches [47] are required at the auxiliary PA output, which either demand extra chip area or degrade reliability. Removing these overhead at the auxiliary PA output leads to Doherty-like operation and degraded PA efficiency [67], [68]. Doherty PAs that use series power combiners together with voltage-mode PAs such as switched-capacitor PAs [69] have been recently developed at sub-6GHz [50]-[52]. They achieve true Doherty load modulation without the overhead of passives or switches at the auxiliary PA output. However, designing efficient voltage-mode PAs at mm-wave entails challenges.

### B. Low-Loss and Broadband Doherty Power Combiner

#### 1) Theoretical Analyses Based On the $\lambda/4$-Line Model

In the field of microwave theory and techniques, a three-$\lambda/4$-line-based Doherty power combiner is presented (Fig. 5) [70]-[74]. It combines power from two current-mode PAs and achieves true Doherty load modulation. In this network, the $\lambda/4$ line at the main PA output, TL1 in Fig. 5, acts as an impedance inverter, and its characteristic impedance is derived as

$$Z_{01} = (1+\alpha)\sqrt{\frac{R_{opt}R_L}{2}} . \quad (5)$$

Meanwhile, the characteristic impedances of the two $\lambda/4$ lines at the auxiliary PA output, TL$_2$ and TL$_3$ in Fig. 5 should satisfy

$$\frac{Z_{03}}{Z_{02}} = \sqrt{\frac{2R_L}{R_{opt}}} . \quad (6)$$

$Z_{02}$ and $Z_{03}$ can be arbitrarily chosen in their physically achievable range, as long as their ratio meets (6). For example, if choosing $Z_{02} = Z_{01} = (1+\alpha)\sqrt{R_{opt}R_L/2}$, $Z_{03} = (1+\alpha)R_L$.

When the auxiliary PA is on, i.e., $i_{main} \in [2/(1+\alpha)^2, 2/(1+\alpha)]$, the ITR of TL$_1$ in Fig. 5 is derived as

$$ITR_{Intro} = \begin{cases} \beta & , if \, \beta \geq 1 \\ 1/\beta, if \, \beta < 1 \end{cases} , \quad (7)$$

where

$$\beta = \frac{R_{opt}}{2R_L} [\frac{1+\alpha}{(2+\alpha) - \frac{2}{1+\alpha}/i_{main}}]^2 = \frac{R_{opt}}{2R_L} ITR_{Conv} . \quad (8)$$

Compared with ITR$_{Conv}$, the additional factor of $R_{opt}/2R_L$ in (8) leads to fundamental changes. Mathematically, $\beta$ is not always $\geq 1$ when the auxiliary PA is on. Therefore, we split (7) into two cases to ensure that ITR$_{Intro} \geq 1$. Physically, this means that the ITR has become dependent on $R_{opt}$ and thus the Doherty PA peak P$_{out}$. To explore the characteristics of this three-$\lambda/4$-line-based Doherty power combiner in more depth, we consider two scenarios. First, in the symmetric design ($\alpha=1$) with a given $R_{opt}$ (and thus peak P$_{out}$), (8) is simplified as

$$\beta(\alpha=1) = \frac{2R_{opt}}{R_L(3-1/i_{main})^2} = \frac{R_{opt}}{2R_L} \cdot ITR_{Conv}(\alpha=1) . \quad (9)$$

In (4), we observe the monotonicity of $ITR_{Conv}$ with respect to $i_{main}$. The factor of $R_{opt}/2R_L$ in (9) can eliminate this monotonicity for $ITR_{Intro}$. For example, $R_{opt}=41.3\Omega$ in our prototype; during PBO, $ITR_{Intro}$ in the symmetric design first decreases from 2.42 at the peak $P_{out}$ ($i_{main}=1$) to unity at 4.7-dB PBO ($i_{main}=0.583$), and then increases to 1.65 at 6-dB PBO ($i_{main}=0.5$) (Fig. 6). Compared with the two-$\lambda/4$-line-based design, the ITR is reduced by 2.42 times at 6-dB PBO while achieving the same peak PA $P_{out}$. Such reduction in the ITR enhances passive efficiency in PBO and bandwidth. Second, considering the ITR at the second efficiency peak, we substitute $i_{main}=2/(1+\alpha)^2$ into (8), and

$$\beta[i_{main} = \frac{2}{(1+\alpha)^2}] = \frac{(1+\alpha)^2 R_{opt}}{2R_L} = \frac{R_{opt}}{2R_L} \cdot ITR_{Conv}[i_{main} = \frac{2}{(1+\alpha)^2}] . \quad (10)$$

Unlike $ITR_{Conv}$, $ITR_{Intro}$ at the second efficiency peak may not monotonically increase with respect to the asymmetric ratio $\alpha$, again due to the additional factor of $R_{opt}/2R_L$. In fact, the asymmetric design technique using a stronger auxiliary PA may be leveraged to minimize the ITR at the second efficiency peak and enhance the passive efficiency. For example, if $R_{opt}<R_L/2$, the auxiliary PA can be designed as ($\sqrt{2R_L/R_{opt}}-1$) times strong as the main PA so that no impedance transformation is required at the second efficiency peak. In other words, we may combine the three-$\lambda/4$-line-based Doherty power combiner with the asymmetric Doherty PA design technique to simultaneously achieve enhanced passive and active efficiencies in deep PBO. Since $R_{opt}=41.3\Omega>R_L/2$ in our prototype, we choose the symmetric design. As per our previous discussion, the three-$\lambda/4$-line-based Doherty power combiner achieves 2.42 times ITR reduction at the second efficiency peak in the symmetric design (Fig. 6). Such ITR reduction enhances passive efficiency in PBO and bandwidth.

### 2) Transformer-Based Low-Loss and Broadband Doherty Power Combiner

The three-$\lambda/4$-line-based Doherty power combiner has been demonstrated in board-level designs [76]-[80]. However, the direct implementation of three $\lambda/4$ lines on-chip can be area consuming. We introduce a new transformer-based solution to achieve this three-$\lambda/4$-line-based low-loss and broadband Doherty power combiner [60]. It only occupies a two-transformer footprint, which is very compact. It also absorbs device parasites, which is a broadband practice.

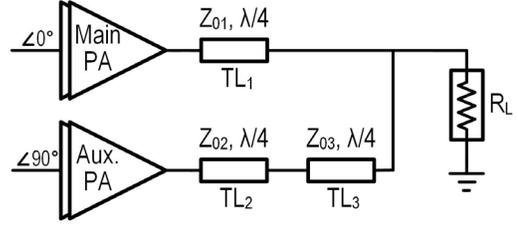

Fig. 5. Three-$\lambda/4$-line-based parallel Doherty power combiner.

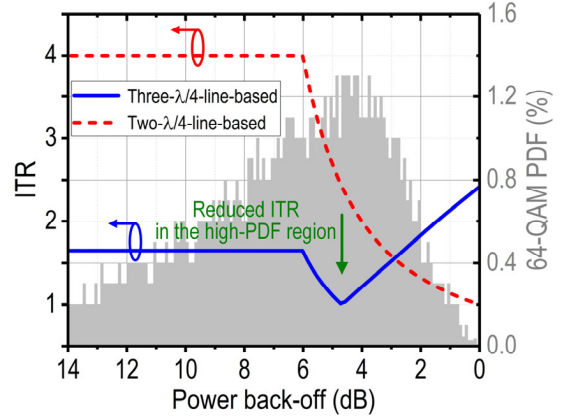

Fig. 6. ITR of the impedance inverter in the three-$\lambda/4$-line-based parallel Doherty power combiner ($R_{opt}=41.3\Omega$).

Next, we describe the process of network synthesis for the new transformer-based Doherty power combiner. First, the three $\lambda/4$ lines in Fig. 5 are approximated using $\pi$-networks (Fig. 7a). $TL_1$ and $TL_2$ are approximated using low-pass $\pi$-networks, and $TL_3$ is approximated using a high-pass $\pi$-network. To compensate the phase responses of the $\pi$-networks and ensure in-phase power combining at the Doherty PA output, the input phases of the two PA paths are offset accordingly (Fig. 7a). In the second step, two ideal transformers are inserted into this network (Fig. 7b). In the third step, the inductors and capacitors in the three $\pi$-networks are reorganized (Fig. 7c). Series inductors are paired with shunt inductors, and four inductors form two groups. Now, we can notice two on-chip transformer models [81]. In other words, series and shunt inductors are absorbed as the leakage and magnetization inductors of two on-chip transformers. As a result, the three-$\lambda/4$-line-based Doherty power combiner is realized in a two-transformer footprint (Fig. 7d). $C_1$ and $C_2$ in Fig. 7d can absorb the parasitic capacitances of PA active cells, and $C_3$ in Fig. 7d can absorb the parasitic capacitances at the Doherty PA output such as pad capacitances.

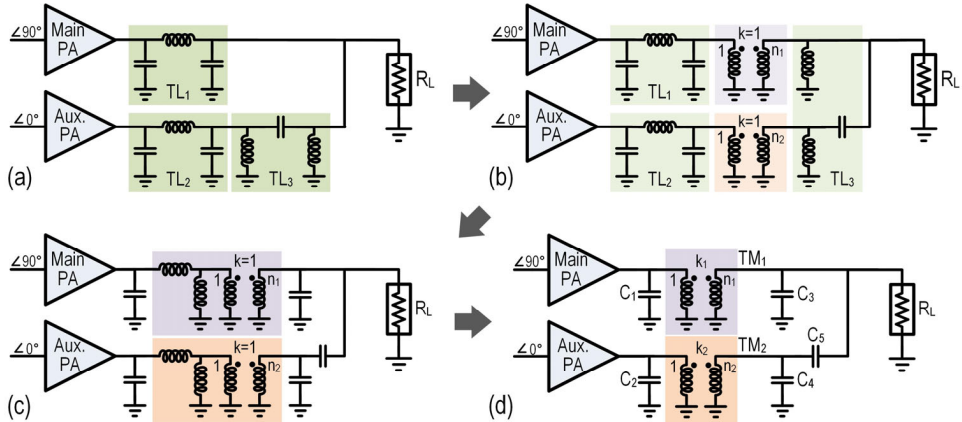

Fig. 7. Introduced transformer-based low-loss and broadband Doherty power combiner.

Based on this process of network synthesis, the closed-form design equations of all the parameters in this network are derived. In the following discussions, we will show the derivations in the symmetric Doherty PA design. Denote the turn ratios and magnetic coupling coefficients of the two transformers as $n_1$ and $k_1$ for $TF_1$ in the main PA path and $n_2$ and $k_2$ for $TF_2$ in the auxiliary PA path. The ideal transformers in the on-chip transformer models bring adjustments to (5) and (6). The characteristic impedance of the CLC low-pass $\pi$-network in the main PA path is now

$$Z_{0\_LP\_main} = \frac{k_1}{n_1}\sqrt{2R_{opt}R_L} \ . \tag{11}$$

The characteristic impedances of the CLC low-pass and LCL high-pass $\pi$-networks in the auxiliary PA path now satisfy

$$\frac{Z_{0\_HP\_aux}}{Z_{0\_LP\_aux}} = \frac{n_2}{k_2}\sqrt{\frac{2R_L}{R_{opt}}} \ . \tag{12}$$

From (11), $L_{p1}$, the primary inductance of $TF_1$, and $C_1$, $C_3$ in Fig. 7d are calculated as

$$L_{p1} = \frac{Z_{0\_LP\_main}}{\omega(1-k_1^2)} = \frac{k_1}{\omega n_1(1-k_1^2)}\sqrt{2R_{opt}R_L} \ , \tag{13}$$

$$C_1 = \frac{1}{\omega Z_{0\_LP\_main}} = \frac{n_1}{\omega k_1\sqrt{2R_{opt}R_L}} \ , \tag{14}$$

and

$$C_3 = (\frac{k_1}{n_1})^2 C_1 \ . \tag{15}$$

Furthermore, we utilize the physical constraint posed by the relationship between the leakage and magnetization inductances in the transformer model. The magnetization inductances in $TF_1$ and $TF_2$, $L_{m1}$ and $L_{m2}$, have

$$L_{m1} = \frac{k_1^2 Z_{0\_LP\_main}}{\omega(1-k_1^2)} = \frac{Z_{0\_HP\_aux}}{\omega(n_1/k_1)^2} \tag{16}$$

and

$$L_{m2} = \frac{k_2^2 Z_{0\_LP\_aux}}{\omega(1-k_2^2)} = \frac{Z_{0\_HP\_aux}}{\omega(n_2/k_2)^2} \ . \tag{17}$$

From (16),

$$Z_{0\_HP\_aux} = \frac{n_1^2}{1-k_1^2} Z_{0\_LP\_main} = \frac{n_1 k_1}{1-k_1^2}\sqrt{2R_{opt}R_L} \ . \tag{18}$$

Then, $L_{p2}$, the primary inductance of $TF_2$, and $C_5$ in Fig. 7d are calculated as

$$L_{p2} = \frac{Z_{0\_HP\_aux}}{\omega(n_2/k_2)^2 k_2^2} = \frac{n_1 k_1}{\omega n_2^2(1-k_1^2)}\sqrt{2R_{opt}R_L} \tag{19}$$

and

$$C_5 = \frac{1}{\omega Z_{0\_HP\_aux}} = \frac{1-k_1^2}{\omega n_1 k_1\sqrt{2R_{opt}R_L}} \ . \tag{20}$$

When $n_1 = n_2$, $L_{p1}$ in (13) equals $L_{p2}$ in (19). In other words, the two transformers have the same primary inductance if their turn ratios are the same. From (17),

$$\frac{Z_{0\_HP\_aux}}{Z_{0\_LP\_aux}} = \frac{n_2^2}{1-k_2^2} \ . \tag{21}$$

Based on (12) and (21), we solve $k_2$ as

$$k_2 = \frac{\sqrt{n_2^2 R_{opt}/(2R_L)+4} - n_2\sqrt{R_{opt}/(2R_L)}}{2} \ . \tag{22}$$

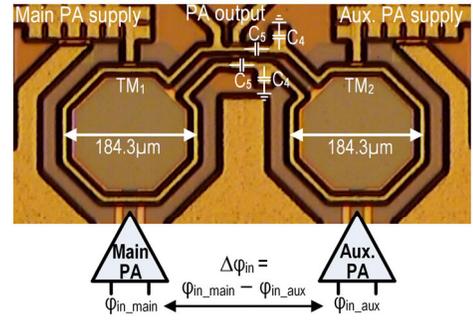

Fig. 8. Implemented transformer-based Doherty power combiner.

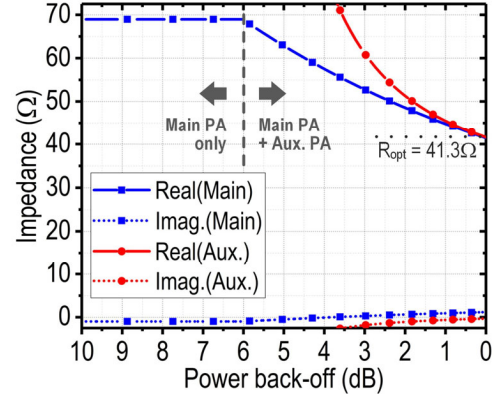

Fig. 9. Simulated effective load impedances based on 3D EM.

The beauty of this result is that $k_2 \in (0,1)$ is always true, which aligns with the physical meaning of $k_2$. Using the solved $k_2$, we calculate

$$C_2 = \frac{1}{\omega Z_{0\_LP\_aux}} = \frac{n_2^2}{\omega(1-k_2^2)Z_{0\_HP\_aux}}$$
$$= \frac{n_2^2(1-k_1^2)}{\omega n_1 k_1(1-k_2^2)\sqrt{2R_{opt}R_L}} \tag{23}$$

and

$$C_4 = (\frac{k_2}{n_2})^2 C_2 \ . \tag{24}$$

In summary, for given load-pull impedance $R_{opt}$, $n_1$, $k_1$, and $n_2$, all the design parameters in the introduced transformer-based Doherty power combiner can be calculated, as (13)-(15), (19), (20), (22)-(24).

Fig. 8 shows our implemented transformer-based Doherty power combiner. It occupies a two-transformer footprint, which is very compact. Fig. 9 shows the simulated effective load impedances seen by the main and auxiliary PAs. The real parts achieve true Doherty load modulation, and the imaginary parts are tuned out for both PAs without using any extra passive matching networks or switches. Fig. 10 compares the simulated passive efficiency and bandwidth of our transformer-based Doherty power combiner with the conventional two-$\lambda$/4-line-based design. We observe substantial passive efficiency improvement in PBO and bandwidth enhancement for the introduced Doherty power combiner.

## III. Power-Dependent Doherty PA Uneven-Feeding Scheme

The introduced transformer-based low-loss and broadband Doherty power combiner addresses the challenge on the passive design in integrated Doherty PAs. On the active side, a two-way Doherty PA demands careful cooperation between the main and auxiliary PAs to ensure the efficiency and linearity. To address this challenge, we introduce a power-dependent Doherty PA uneven-feeding scheme [60].

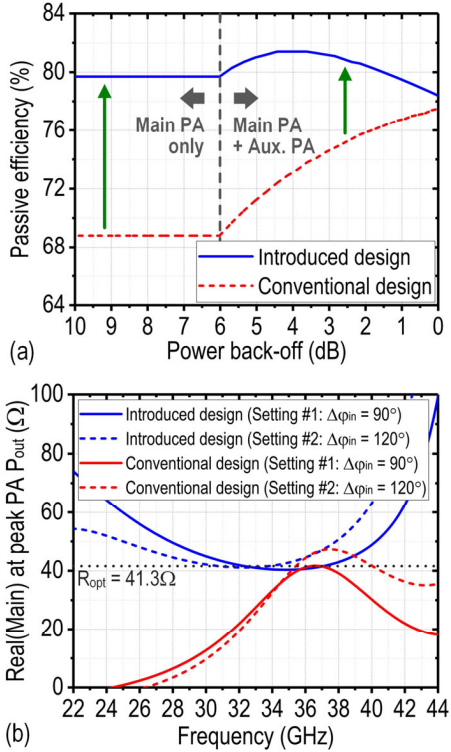

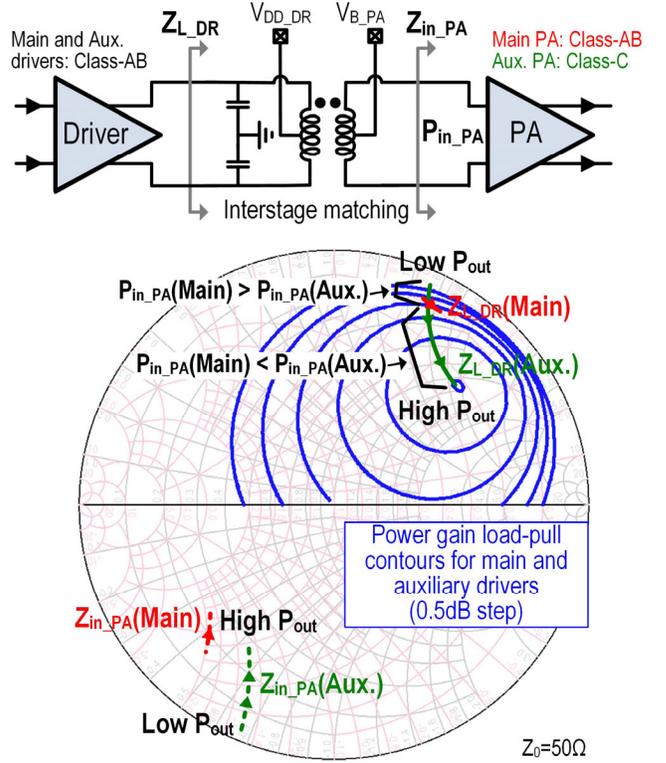

Fig. 10. Simulated (a) passive efficiency and (b) bandwidth of our introduced Doherty power combiner and their comparisons with a conventional two-$\lambda$/4-line-based design.

Fig. 11. Introduced power-dependent Doherty PA uneven-feeding scheme.

Fig. 2b shows the desired current relationship between the main and auxiliary PAs in the ideal Doherty operation. The auxiliary PA should be off in the low-power region, and after it is turned on in the high-power region, its current need to increase rapidly. To achieve the late turning-on of the auxiliary PA, the analog Doherty PA often biases the auxiliary PA in class-C and the main PA in class-AB. To achieve the rapidly increasing current from the auxiliary PA, the conventional analog Doherty PA often adopts dynamic biasing that provides higher biasing levels for the auxiliary PA as the PA input power ($P_{in}$) increases [65]. However, the dynamic biasing circuit is loaded by the large power cells, and it need to track the envelope, which has three to five times bandwidth expansion on top of the input modulated signal. As a result, conventional Doherty PAs using the dynamic biasing technique entail challenges for broadband 5G applications.

Our introduced power-dependent Doherty PA uneven-feeding scheme facilitates the rapid increase of the auxiliary PA current and achieves enhanced Doherty operation without hardware overhead or bandwidth sacrifice. Fundamentally, the introduced scheme leverages the different $P_{in}$ dependences of the input impedances in the differently biased main and auxiliary PAs using bipolar transistors. When the PA $P_{in}$ increases, the class-C-biased auxiliary PA is gradually turned on, and its effective transconductance increases. As a result, its input conductance increases significantly (dashed green line in Fig. 11). On the other hand, the input impedance of the class-AB-biased main PA remain almost the same when the PA $P_{in}$ changes (dashed red line in Fig. 11). We introduce a "driver-PA co-design" method that leverages the different $P_{in}$ dependences and creates adaptive uneven feeding for the main and auxiliary PA final stages (Fig. 11). The main and auxiliary drivers share the same biasing, and the blue curves in Fig. 11 show their power-gain load-pull contours. Solid red and green lines in Fig. 11 show the load impedances seen by the main and auxiliary drivers, respectively, which are transformed from the input impedances of the main and auxiliary PAs by the inter-stage matching networks. When the $P_{in}$ increases, the load impedance trajectory of the auxiliary driver travels from the low-power-gain region to the high-power-gain region, while the load impedance trajectory of the main driver almost stays on a constant power-gain circle. As a result, the auxiliary PA final stage is fed by an expanding $P_{in}$ that generates its rapidly increasing output current in the high-power region. Existing uneven-feeding techniques achieve uneven power division at the Doherty input, which is coupled with the input matching and limits the design freedom [82]-[85]. We achieve power-dependent adaptive uneven feeding by the introduced "driver-PA co-design" method without hardware overhead and sacrificing modulation bandwidth.

## IV. EXPERIMENTAL RESULTS

### A. PA Implementation

For the proof of concept, we implement a 28/37/39-GHz multiband Doherty PA for broadband 5G applications (Fig. 12). It is prototyped in a GlobalFoundries 130-nm SiGe BiCMOS process and occupies 1.8mm² (Fig. 13).

The input signal is first processed by an on-chip transformer-based differential quadrature hybrid [86], which splits the input power and performs 90° phase shifts. The quadrature hybrid is loaded by nine-section varactor-loaded transmission lines. The varactor control voltages in the main and auxiliary paths can be independently tuned, which adjusts the relative phase between the two paths [55], [87] and extends the carrier bandwidth of a Doherty PA [55], [87]. Our design uses one varactor setting for 28GHz and the other varactor setting for 37/39GHz (Table I). The varactor-loaded transmission lines also form high-order loads for the quadrature hybrid, which ensures wideband input matching for different varactor settings.

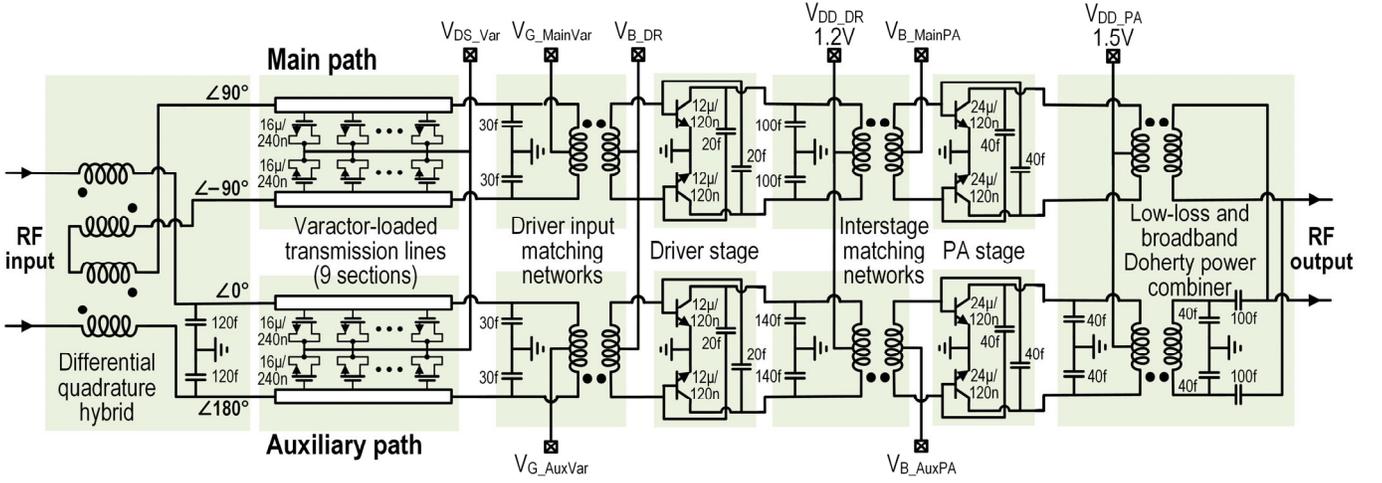

Fig. 12. Schematic of the PA prototype.

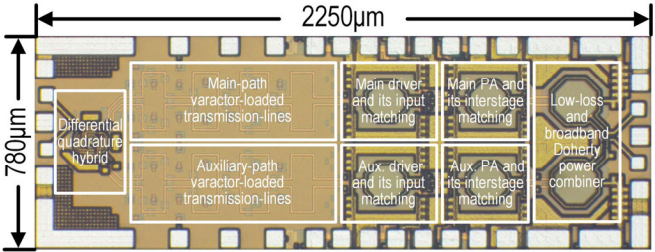

Fig. 13. Chip microphotograph.

Table I
Varactor settings for target band(s)

|  | Target band(s) | $V_{DS\_Var} - V_{G\_MainVar}$ | $V_{DS\_Var} - V_{G\_AuxVar}$ |
|---|---|---|---|
| Setting #1 | 28GHz | 0.5 | 0 |
| Setting #2 | 37GHz, 39GHz | 0 | 0.5 |

The PA is comprised of two stages. Both stages adopt the neutralization technique that enhances the power gain and stability [19]. The driver stage is connected to the varactor-loaded transmission lines with transformer-based matching. The inter-stage matching between the driver and PA final stage is designed to achieve the introduced power-dependent uneven-feeding scheme.

### B. Measurement Results

#### 1) Continuous-Wave (CW) Measurement

The PA is first characterized using CW signals. Fig. 14 and Fig. 15 show the measured small-signal S-parameters and large-signal saturated $P_{out}$ ($P_{sat}$) and $P_{1dB}$, which demonstrate broadband performance. The small-signal $S_{21}$ achieves a $-3$-dB bandwidth of 23.3-39.7GHz (52% fractional bandwidth). The $-1$-dB $P_{sat}$ bandwidth covers 28 to 42GHz (40% fractional bandwidth).

Fig. 16 shows measured PBO performance. At 37GHz, this PA achieves +17.1-dBm $P_{sat}$, +15.5-dBm $P_{1dB}$, 27.6% peak collector efficiency (CE), and 22.6% peak power added efficiency (PAE). Compared with a normalized class-B/class-A PA, the Doherty operation achieves 1.92/3.86 times efficiency enhancement at 6-dB PBO. The PA also demonstrates excellent amplitude and phase linearity at 37GHz. The AM-PM is 1.3° from the small signal to $P_{1dB}$. Without changing the varactor controls, the PA achieves +17-dBm $P_{sat}$, +15.4-dBm $P_{1dB}$, 28.2% peak CE, and 21.4% peak PAE at 39GHz. The Doherty efficiency enhancement at 6.7-dB PBO is 1.62/3.51 times over the class-B/class-A operation. After changing the band setting by reconfiguring the varactor controls (Table I), the PA achieves +16.8-dBm $P_{sat}$, +15.2-dBm $P_{1dB}$, 29.4% peak CE, and 20.3% peak PAE at 28GHz. The Doherty efficiency enhancement at 5.9-dB PBO is 1.72/3.39 times over the class-B/class-A operation. Owing to the introduced Doherty power combiner and adaptive uneven feeding scheme, the PA achieves superior PBO efficiency improvement in all the three 5G bands.

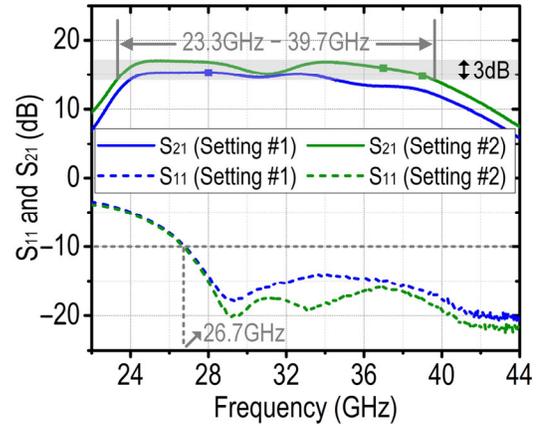

Fig. 14. Measured small-signal S-parameters.

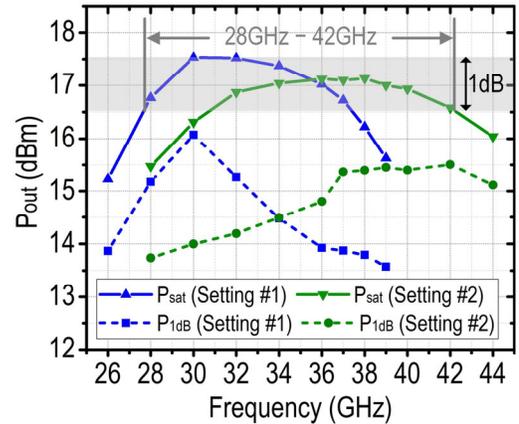

Fig. 15. Measured large-signal $P_{sat}$ and $P_{1dB}$.

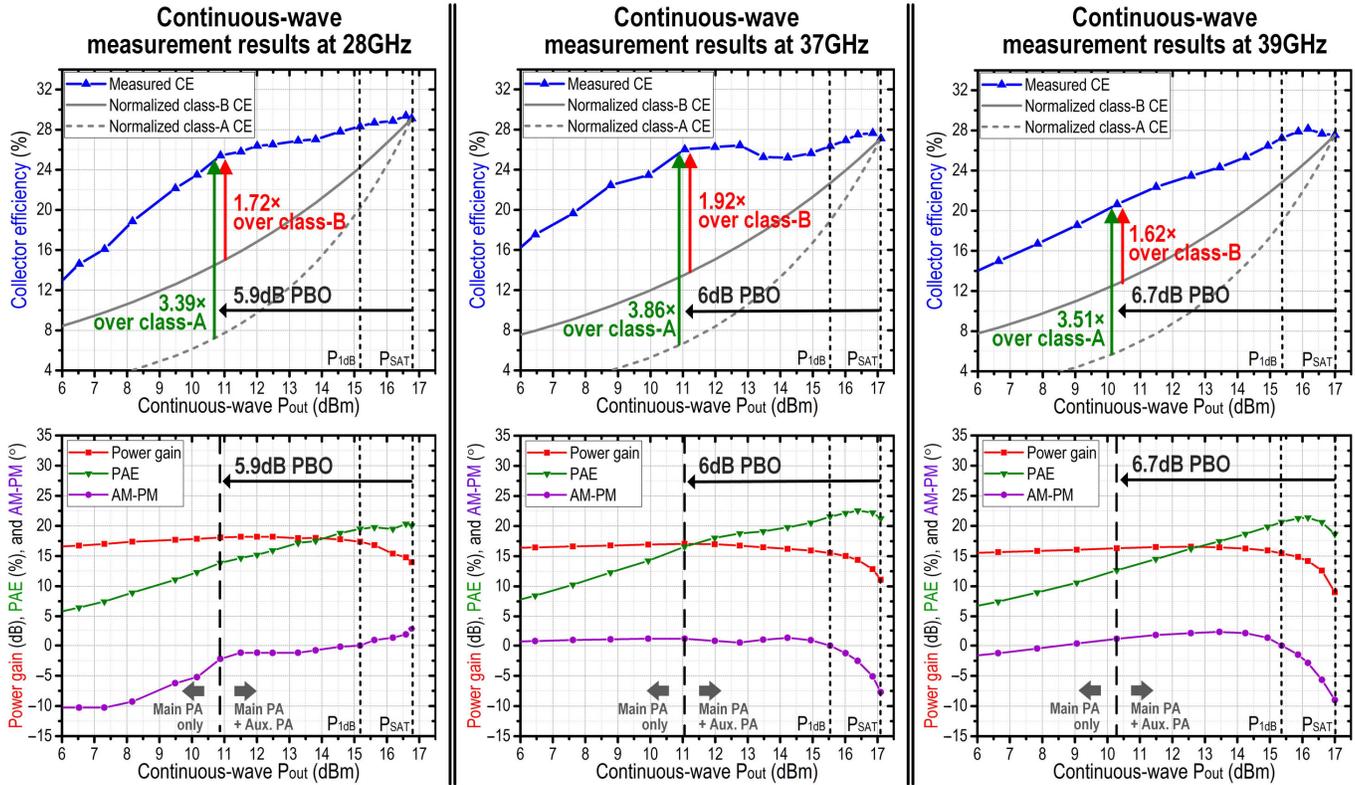

Fig. 16. Measured PBO performance at 28GHz, 37GHz, and 39GHz.

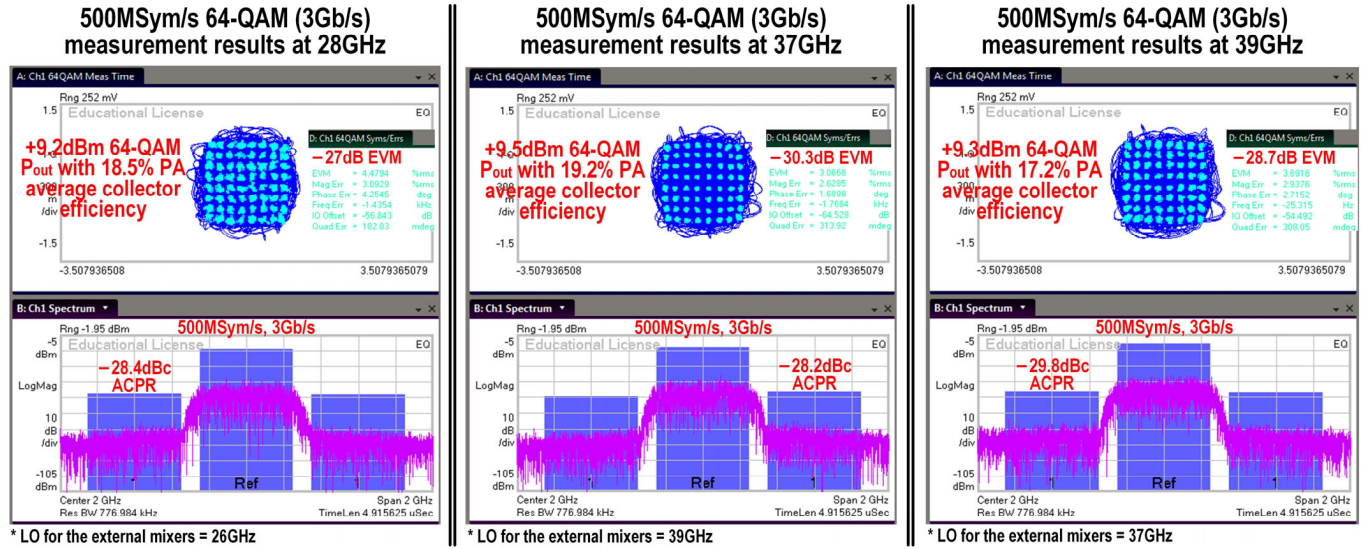

Fig. 17. Measurement results for 500M-Sym/s 64QAM (3Gb/s) at 28GHz, 37GHz, and 39GHz.

### 2) Modulation Measurement

The PA is then characterized using modulated signals. In the measurements, we calibrate out the non-ideal effects due to cables, off-chip components, and instruments in the setup via the equalization function in the vector signal analysis (VSA) software [88]. The coefficients of the equalization filter are determined by the through tests without our PA.

Fig. 17 shows the measurement results using 500M-Sym/s 64QAM (3Gb/s). In the 37/39/28-GHz 5G bands, the PA delivers +9.5/+9.3/+9.2-dBm average $P_{out}$, 19.2/17.2/18.5% average CE, −30.3/−28.7/−27-dB rms error vector magnitude (EVM), and −28.2/−29.8/−28.4-dBc adjacent channel power ratio (ACPR). Fig. 18 shows the measurement results using 1G-Sym/s 64QAM (6Gb/s) at 28GHz. The PA achieves +7.2-dBm average $P_{out}$, −26.6-dB rms EVM, and −25.4-dBc ACPR. These measurements verify the multiband linear Doherty performance in high-speed dynamic operations.

Table II summarizes the performance of our PA and makes comparisons with state-of-the-art mm-wave PAs in silicon.

## V. Conclusion

We present the first multiband mm-wave linear Doherty PA in silicon for 5G applications. To address the unmet challenges in conventional designs, we introduce a transformer-based low-loss and broadband on-chip Doherty power combiner and a power-dependent Doherty PA uneven-feeding scheme based on a "driver-PA co-design" method. Our prototype demonstrates multi-Gb/s data rates in three mm-wave 5G bands with excellent efficiency and linearity. Our PA advances the state of the art for Doherty, wideband, and 5G silicon PAs in mm-wave bands. Moreover, our PA allows the drop-in upgrade for current PAs in existing mm-wave systems and opens doors to compact system solutions for future multiband 5G massive MIMO and phased-array platforms.

## Acknowledgement

The authors would like to acknowledge GlobalFoundries for foundry service.

## References

[1] www.3gpp.org.

[2] A. Seam, *et al.*, "Enabling mobile augmented and virtual reality with 5G networks," AT&T, Jan. 2017.

[3] "Augmented and virtual reality: the first wave of 5G killer apps," ABI Research and Qualcomm, Feb. 2017.

[4] T. S. Rappaport, *et al.*, "Millimeter wave mobile communications for 5G cellular: It will work!" *IEEE Access*, vol. 1, pp. 335–349, May 2013.

[5] S. Rangan, T. Rappaport, and E. Erkip, "Millimeter-wave cellular wireless networks: potentials and challenges," *Proc. IEEE*, vol. 102, no. 3, pp. 366–385, Mar. 2014.

[6] https://apps.fcc.gov/edocs_public/attachmatch/FCC-16-89A1.pdf

[7] "Spectrum for 4G and 5G", Qualcomm, Dec. 2017.

[8] C. Balanis, *Antenna Theory: Analysis and Design*, 4th ed. Hoboken, NJ: Wiley, 2016.

[9] W. Zhai, V. Miraftab, M. Repeta, D. Wessel, and W. Tong, "Dual-band millimeter-wave interleaved antenna array exploiting low-cost PCB technology for high speed 5G communication," in *Proc. IEEE Int. Microw. Symp.*, 2016, pp. 1–4.

[10] S. Jilani and A. Alomainy, "A multiband millimeter-wave 2D array based on enhanced franklin antenna for 5G wireless

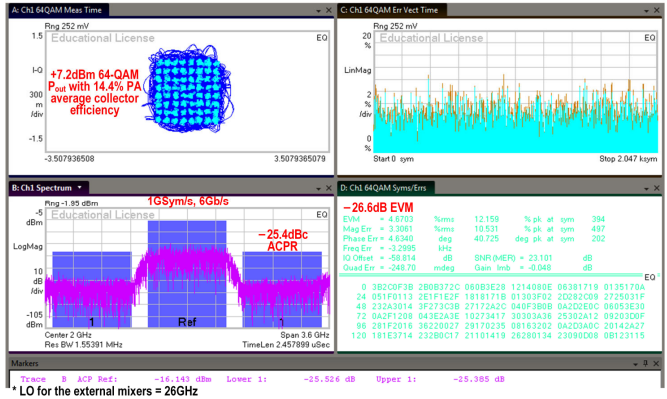

Fig. 18. Measurement results for 1G-Sym/s 64QAM (6Gb/s) at 28GHz.

Table II
Performance summary and comparisons

| | This work | | | Mm-wave Doherty PA | | Mm-wave wideband PA | | | | 5G PA | |
| --- | --- | --- | --- | --- | --- | --- | --- | --- | --- | --- | --- |
| | | | | [27] A. Agah, JSSC '13 | [29] E. Kaymaksut, TMTT '15 | [30] M. Bassi, JSSC '15 | [37] C. Chappidi, JSSC '17 | [43] M. Vigilante, JSSC '18 | [35] S. Shakib, JSSC '16 | [38] A. Sarkar, JSSC '17 |
| Technology | 130nm SiGe | | | 45nm SOI CMOS | 40nm CMOS | 28nm CMOS | 130nm SiGe | 28nm CMOS | 28nm CMOS | 180nm SiGe |
| Architecture | Multiband Doherty | | | Slow-wave CPW Doherty† | Doherty | Inductively coupled resonator | Asymmetric combiner | Class-AB with AM-PM compensation | Inductive source degeneration | Four-way power combined |
| Supply (V) | 1.5 | | | 2.5 | 1.5 | 1 | 4 | 0.9 | 1 | 3.8 |
| S$_{21}$ −3dB BW (GHz) | 23.3–39.7 (52%) | | | N.A. | 60–81 (30%) | 40–67 (51%) | 32–60 (61%)* | 29–57 (65%) | 27–31 (14%)* | 26–35 (30%)* |
| P$_{sat}$ −1dB BW (GHz) | 28–42 (40%) | | | N.A. | 57–77 (28%) | 46–62 (30%) | 40–65 (48%) | 56.6% | 28–31 (10%)* | N.A. |
| Area (mm²) | 1.76 | | | 0.64 | 0.96 | 0.33 | 1.02 | 0.16 | 0.16** | 0.71 |
| Frequency (GHz) | 28 | 37 | 39 | 42 | 72 | 53 | 55 | 30 | 30 | 28 |
| Power gain (dB) | 18.2 | 17.1 | 16.6 | 7 | 18.6* | 13 | 18.8 | 18.3* | 15.7 | 28.6 |
| P$_{sat}$ (dBm) | +16.8 | +17.1 | +17 | +18 | +21 | +13.3 | +23.6 | +16.6 | +14 | +23.7 |
| P$_{1dB}$ (dBm) | +15.2 | +15.5 | +15.4 | +14.5* | +19.2 | +12 | +19.9 | +13.4 | +13.2 | +23.2 |
| Peak η | 29.4% CE 20.3% PAE | 27.6% CE 22.6% PAE | 28.2% CE 21.4% PAE | 33% DE 23% PAE | 20.7% DE 13.6% PAE | 16% PAE | 27.7% PAE | 24.2% PAE | 35.5% PAE | 32.7% PAE |
| η @ P$_{1dB}$ | 28.3% CE 19.5% PAE | 26.4% CE 21.6% PAE | 27.2% CE 20.7% PAE | 26% DE* 19% PAE* | 17.6% DE* 12.4% PAE | 14% PAE | 15% PAE | 12.6% PAE | 34.3% PAE | 32.7% PAE |
| η @ PBO | 25.4% CE 13.9% PAE @5.9dB PBO | 26% CE 16.6% PAE @6dB PBO | 20.6% CE 12.6% PAE @6.7dB PBO | 24% DE 17% PAE @6dB PBO | 11.5% DE* 7% PAE @6dB PBO | 5% PAE* @6dB PBO | 7% PAE* @6dB PBO | 7.5% PAE* @6dB PBO | 10% PAE @9.6dB PBO | 15% PAE @6dB PBO |
| Modulation results | 64-QAM 3Gb/s +9.2dBm −27dB EVM −28.44dBc ACPR 18.5% CE | 64-QAM 3Gb/s +9.5dBm −30.3dB EVM −28.28dBc ACPR 19.2% CE | 64-QAM 3Gb/s +9.4dBm −28.7dB EVM −29.8dBc ACPR 17.2% CE | N.A. | 64-QAM 0.6Gb/s +15.9dBm −25.6dB EVM 7.2% PAE | N.A. | 64-QAM 3Gb/s +12.8dBm −25.5dB EVM | 64-QAM 3Gb/s +4.6dBm* <−25dB EVM −41.3dBc ACPR* 1.7% PAE* | 64-QAM 1.5Gb/s +4.2dBm −25dB EVM −26.4dBc ACPR 9% PAE | N.A. |
| | 64-QAM 6Gb/s +7.2dBm −26.6dB EVM −25.4dBc ACPR 14.4% CE | | | | | | | 64-QAM 6Gb/s +3.6dBm* <−25dB EVM −40.2dBc ACPR* 1.3% PAE* | | |

* Read from the reported figures. ** Without pads. † Statically tuned biasing.

systems," *IEEE Antennas Wireless Propag. Lett.*, vol. 16, pp. 2983–2986, 2017.

[11] L. Li, "IC challenges in 5G," in *Proc. IEEE Asian Solid-State Circuits Conf.*, 2015, pp. 1–4.

[12] S. Cripps, *RF Power Amplifiers for Wireless Communications*, 2nd ed. Boston, MA, USA: Artech House, 2006.

[13] F. Raab, *et al.*, "Power amplifiers and transmitters for RF and microwave," *IEEE Trans. Microw. Theory Techn.*, vol. 50, no. 3, pp. 814–826, Mar. 2002.

[14] H. Wang and K. Sengupta, *RF and mm-Wave Power Generation in Silicon*. Cambridge, MA, USA: Academic Press, 2015.

[15] H. Hashemi and S. Raman, *mm-Wave Silicon Power Amplifiers and Transmitters*. Cambridge, UK: Cambridge University Press, 2016.

[16] A. Komijani and A. Hajimiri, "A wideband 77-GHz, 17.5-dBm power amplifier in silicon," in *Proc. IEEE Custom Integrated Circuits Conf.*, 2005, pp. 571–574.

[17] A. Arbabian and A. Niknejad, "A broadband distributed amplifier with internal feedback providing 660-GHz GBW in 90-nm CMOS," in *IEEE Int. Solid-State Circuits Conf. Tech. Dig.*, 2008, pp. 196–197.

[18] J. Lai and A. Valdes-Garcia, "A 1-V 17.9-dBm 60-GHz power amplifier in standard 65-nm CMOS," in *IEEE Int. Solid-State Circuits Conf. Tech. Dig.*, 2010, pp. 424–425.

[19] W. Chan and J. Long, "A 58-65 GHz neutralized CMOS power amplifier with PAE above 10% at 1-V supply," *IEEE J. Solid-State Circuits*, vol. 45, no. 3, pp. 554–564, Mar. 2010.

[20] J. Chen and A. Niknejad, "A compact 1-V 18.6-dBm 60-GHz power amplifier in 65-nm CMOS," in *IEEE Int. Solid-State Circuits Conf. Tech. Dig.*, 2011, pp. 432–433.

[21] Q. Gu, Z. Xu, and F. Chang, "Two-way current-combining W-band power amplifier in 65-nm CMOS," *IEEE Trans. Microw. Theory Techn.*, vol. 60, no. 5, pp. 1365–1374, May 2012.

[22] Y. Zhao and J. Long, "A wideband, dual-path, millimeter-wave power amplifier with 20-dBm output power and PAE above 15% in 130-nm SiGe BiCMOS," *IEEE J. Solid-State Circuits*, vol. 47, no. 9, pp. 1981–1997, Sept. 2012.

[23] D. Zhao, S. Kulkarni, and P. Reynaert, "A 60-GHz outphasing transmitter in 40-nm CMOS," *IEEE J. Solid-State Circuits*, vol. 47, no. 12, pp. 3172–3183, Dec. 2012.

[24] H. Dabag, B. Hanafi, F. Golcuk, A. Agah, J. Buckwalter, and P. Asbeck, "Analysis and design of stacked-FET millimeter-wave power amplifiers," *IEEE Trans. Microw. Theory Techn.*, vol. 61, no. 4, pp. 1543–1556, Apr. 2013.

[25] A. Balteanu *et al.*, "A 2-bit, 24-dBm, millimeter-wave SOI CMOS power-DAC cell for watt-level high-efficiency, fully digital m-ary QAM transmitters," *IEEE J. Solid-State Circuits*, vol. 48, no. 5, pp. 1126–1137, May 2013.

[26] D. Zhao and P. Reynaert, "A 60-GHz dual-mode class-AB power amplifier in 40-nm CMOS," *IEEE J. Solid-State Circuits*, vol. 48, no. 10, pp. 2323–2337, Oct. 2013.

[27] A. Agah, H. Dabag, B. Hanafi, P. Asbeck, J. Buckwalter, and L. Larson, "Active millimeter-wave phase-shift Doherty power amplifier in 45-nm SOI CMOS," *IEEE J. Solid-State Circuits*, vol. 48, no. 10, pp. 2338–2350, Oct. 2013.

[28] A. Chakrabarti and H. Krishnaswamy, "High-power high-efficiency class-E-like stacked mm-wave PAs in SOI and bulk CMOS: theory and implementation," *IEEE Trans. Microw. Theory Techn.*, vol. 62, no. 8, pp. 1686–1704, Aug. 2014.

[29] E. Kaymaksut, D. Zhao, and P. Reynaert, "Transformer-based Doherty power amplifiers for mm-wave applications in 40-nm CMOS," *IEEE Trans. Microw. Theory Techn.*, vol. 63, no. 4, pp. 1186–1192, Apr. 2015.

[30] M. Bassi, J. Zhao, A. Bevilacqua, A. Ghiloni, A. Mazzanti, and F. Svelto, "A 40–67 GHz power amplifier with 13-dBm and 16% PAE in 28-nm CMOS LP," *IEEE J. Solid-State Circuits*, vol. 50, no. 7, pp. 1618–1628, Jul. 2015.

[31] H. Lin and G. Rebeiz, "A 70-80-GHz SiGe amplifier with peak output power of 27.3dBm," *IEEE Trans. Microw. Theory Techn.*, vol. 64, no. 7, pp. 2039–2049, Jul. 2016.

[32] K. Fang, C. Levy, and J. Buckwalter, "Supply-scaling for efficiency enhancement in distributed power amplifiers," *IEEE J. Solid-State Circuits*, vol. 51, no. 9, pp. 1994–2005, Sept. 2016.

[33] S. Mortazavi and K. Koh, "Integrated inverse class-F silicon power amplifiers for high power efficiency at microwave and mm-wave," *IEEE J. Solid-State Circuits*, vol. 51, no. 10, pp. 2420–2434, Oct. 2016.

[34] B. Park *et al.*, "Highly linear mm-wave CMOS power amplifier," *IEEE Trans. Microw. Theory Techn.*, vol. 64, no. 12, pp. 4535–4544, Dec. 2016.

[35] S. Shakib, H. Park, J. Dunworth, V. Aparin, and K. Entesari, "A highly efficient and linear power amplifier for 28-GHz 5G phased array radios in 28-nm CMOS," *IEEE J. Solid-State Circuits*, vol. 51, no. 12, pp. 3020–3036, Dec. 2016.

[36] K. Datta and H. Hashemi, "Watt-level mm-wave power amplification with dynamic load modulation in a SiGe HBT digital power amplifier," *IEEE J. Solid-State Circuits*, vol. 52, no. 2, pp. 371–388, Feb. 2017.

[37] C. Chappidi and K. Sengupta, "Frequency reconfigurable mm-wave power amplifier with active impedance synthesis in an asymmetrical non-isolated combiner: analysis and design," *IEEE J. Solid-State Circuits*, vol. 52, no. 8, pp. 1990–2008, Aug. 2017.

[38] A. Sarkar, F. Aryanfar, and B. Floyd, "A 28-GHz SiGe BiCMOS PA with 32% efficiency and 23-dBm output power," *IEEE J. Solid-State Circuits*, vol. 52, no. 6, pp. 1680–1686, Jun. 2017.

[39] C. Li *et al.*, "5G mm-wave front-end-module design with advanced SOI process," in *Proc. Int. Conf. ASIC*, 2017, pp. 1017–1020.

[40] H. Wang *et al.*, "Towards energy-efficient 5G mm-wave links," in *Proc. IEEE Bipolar/BiCMOS Circuits Technol. Meeting*, 2017, pp. 30–37.

[41] S. Ali, P. Agarwal, J. Baylon, S. Gopal, L. Renaud, and D. Heo, "A 28-GHz 41%-PAE linear CMOS power amplifier using a transformer-based AM-PM distortion-correction technique for 5G phased arrays," in *IEEE Int. Solid-State Circuits Conf. Tech. Dig.*, 2018, pp. 406–408.

[42] T. Li, M. Huang, and H. Wang, "A continuous-mode harmonically tuned 19-to-29.5GHz ultra-linear PA supporting 18Gb/s at 18.4% modulation PAE and 43.5% peak PAE," in *IEEE Int. Solid-State Circuits Conf. Tech. Dig.*, 2018, pp. 410–412.

[43] M. Vigilante and P. Reynaert, "A wideband class-AB power amplifier with 29-57-GHz AM-PM compensation in 0.9-

V 28-nm bulk CMOS," *IEEE J. Solid-State Circuits*, vol. 53, no. 5, pp. 1288–1301, May 2018.

[44] E. McCune, *Dynamic Power Supply Transmitters: Envelope Tracking, Direct Polar, and Hybrid Combinations*. Cambridge, UK: Cambridge University Press, 2015.

[45] D. Chowdhury, S. Mundlapudi, and A. Afsahi, "A fully integrated reconfigurable wideband envelope-tracking SoC for high-bandwidth WLAN applications in a 28-nm CMOS technology," in *IEEE Int. Solid-State Circuits Conf. Tech. Dig.*, 2017, pp. 34–35.

[46] W. H. Doherty, "A new high efficiency power amplifier for modulated waves," *Proc. IRE*, vol. 24, no. 9, pp. 1163–1182, Sept. 1936.

[47] F. H. Raab, "Efficiency of Doherty RF power-amplifier systems," *IEEE Trans. Broadcast.*, vol. BC-33, no. 3, pp. 77–83, Sept. 1987.

[48] A. Grebennikov and S. Bulja, "High-efficiency Doherty power amplifiers: historical aspect and modern trends," *Proc. IEEE*, vol. 100, no. 12, pp. 3190–3219, Dec. 2012.

[49] V. Camarchia, M. Pirola, R. Quaglia, S. Jee, Y. Cho, and B. Kim, "The Doherty power amplifier: Review of recent solutions and trends," *IEEE Trans. Microw. Theory Techn.*, vol. 63, no. 2, pp. 559–571, Feb. 2015.

[50] R. Pengelly, C. Fager, and M. Ozen, "Doherty's legacy: A history of the Doherty power amplifier from 1936 to the present day," *IEEE Microw. Mag.*, vol. 17, no. 2, pp. 41–58, Feb. 2016.

[51] P. Asbeck, "Will Doherty continue to rule for 5G?," in *Proc. IEEE Int. Microw. Symp.*, 2016, pp. 1–4.

[52] R. Darraji, F. Ghannouchi, and O. Hammi, "A dual-input digitally driven Doherty amplifier architecture for performance enhancement of Doherty transmitters," *IEEE Trans. Microw. Theory Techn.*, vol. 59, no. 5, pp. 1284–1293, May 2011.

[53] W. Gaber, P. Wambacq, J. Craninckx, and M. Ingels, "A CMOS IQ digital Doherty transmitter using modulated tuning capacitors," in *Proc. IEEE European Solid State Circuits Conf.*, 2012, pp. 341–344.

[54] S. Hu, S. Kousai, J. Park, O. Chlieh, and H. Wang, "Design of a transformer-based reconfigurable digital polar Doherty power amplifier fully integrated in bulk CMOS," *IEEE J. Solid-State Circuits*, vol. 50, no. 5, pp. 1094–1106, May 2015.

[55] S. Hu, S. Kousai, and H. Wang, "A broadband mixed-signal CMOS power amplifier with a hybrid class-G Doherty efficiency enhancement technique," *IEEE J. Solid-State Circuits*, vol. 51, no. 3, pp. 598–613, Mar. 2016.

[56] V. Vorapipat, C. Levy, and P. Asbeck, "Voltage mode Doherty power amplifier," *IEEE J. Solid-State Circuits*, vol. 52, no. 5, pp. 1295–1304, May 2017.

[57] L. Salem, J. Buckwalter, and P. Mercier, "A recursive switched-capacitor house-of-cards power amplifier," *IEEE J. Solid-State Circuits*, vol. 52, no. 7, pp. 1719–1738, Jul. 2017.

[58] V. Vorapipat, C. Levy, and P. Asbeck, "A class-G voltage-mode Doherty power amplifier," *IEEE J. Solid-State Circuits*, vol. 52, no. 12, pp. 3348–3360, Dec. 2017.

[59] Y. Yin, L. Xiong, Y. Zhu, B. Chen, H. Min, and H. Xu, "A compact dual-band digital Doherty power amplifier using parallel-combining transformer for cellular NB-IoT applications," in *IEEE Int. Solid-State Circuits Conf. Tech. Dig.*, 2018, pp. 408–410.

[60] S. Hu, F. Wang, and H. Wang, "A 28GHz/37GHz/39GHz multiband linear Doherty power amplifier for 5G massive

MIMO applications," in *IEEE Int. Solid-State Circuits Conf. Tech. Dig.*, 2017, pp. 32–33.

[61] S. Hu, S. Kousai, and H. Wang, "Antenna impedance variation compensation by exploiting a digital Doherty power amplifier architecture," *IEEE Trans. Microw. Theory Techn.*, vol. 63, no. 2, pp. 580–597, Feb. 2015.

[62] M. Elmala, J. Paramesh, and R. Bishop, "A 90-nm CMOS Doherty power amplifier with minimum AM-PM distortion," *IEEE J. Solid-State Circuits*, vol. 41, no. 6, pp. 1323–1332, Jun. 2006.

[63] N. Wongkomet, L. Tee, and P. Gray, "A +31.5-dBm CMOS RF Doherty power amplifier for wireless communications," *IEEE J. Solid-State Circuits*, vol. 41, no. 12, pp. 2852–2859, Dec. 2006.

[64] K. Onizuka, S. Saigusa, and S. Otaka, "A +30.5 dBm CMOS Doherty power amplifier with reliability enhancement technique," in *Proc. IEEE Symp. VLSI Circuits*, 2012, pp. 78–79.

[65] K. Onizuka, K. Ikeuchi, S. Saigusa, and S. Otaka, "A 2.4-GHz CMOS Doherty power amplifier with dynamic biasing scheme," in *Proc. IEEE Asian Solid-State Circuits Conf.*, 2012, pp. 93–96.

[66] H. Bode, *Network Analysis and Feedback Amplifier Design*. New York, NY, USA: D. Van Nostrand Company, Inc., 1945.

[67] R. Fano, "Theoretical limitations on the broadband matching of arbitrary impedances," *J. Franklin Inst.*, vol. 249, pp. 57–83, Jan. 1950; pp. 139–154, Feb. 1950.

[68] R. Gilmore and L. Besser, *Practical RF Circuit Design for Modern Wireless Systems*. Boston, MA, USA: Artech House, 2003.

[69] T. Lee, *Planar Microwave Engineering: A Practical Guide to Theory, Measurement, and Circuits*. Cambridge, UK: Cambridge University Press, 2004.

[70] D. Pozar, *Microwave Engineering*, 4th ed. New York, NY, USA: Wiley, 2011.

[71] C. Zhao, B. Park, and B. Kim, "Complementary metal-oxide semiconductor Doherty power amplifier based on voltage combining method," *IET Microw. Antennas Propag.*, vol. 8, no. 3, pp. 131–136, Feb. 2014.

[72] Y. Cho, K. Moon, B. Park, J. Kim, and B. Kim, "Voltage-combined CMOS Doherty power amplifier based on transformer," *IEEE Trans. Microw. Theory Techn.*, vol. 64, no. 11, pp. 3612–3622, Nov. 2016.

[73] E. Kaymaksut and P. Reynaert, "Transformer-based uneven Doherty power amplifier in 90-nm CMOS for WLAN Applications," *IEEE J. Solid-State Circuits*, vol. 47, no. 7, pp. 1659–1671, Jul. 2012.

[74] E. Kaymaksut, B. Francois, and P. Reynaert, "Analysis and optimization of transformer-based power combining for back-off efficiency enhancement," *IEEE Trans. Circuits Syst. I: Reg. Papers*, vol. 60, no. 4, pp. 825–835, Apr. 2013.

[75] S. Yoo, J. Walling, E. Woo, B. Jann, and D. Allstot, "A switched-capacitor RF power amplifier," *IEEE J. Solid-State Circuits*, vol. 46, no. 12, pp. 2977–2987, Dec. 2011.

[76] A. Grebennikov and J. Wong, "A dual-band parallel Doherty power amplifier for wireless applications," *IEEE Trans. Microw. Theory Techn.*, vol. 60, no. 10, pp. 3214–3222, Oct. 2012.

[77] D. Wu, J. Annes, M. Bokatius, P. Hart, E. Krvavac, and G. Tucker, "A 350-W, 790 to 960 MHz wideband LDMOS Doherty amplifier using a modified combining scheme," in *Proc. IEEE Int. Microw. Symp.*, 2014, pp. 1–4.

[78] R. Giofre, L. Piazzon, P. Colantonio, and F. Giannini, "A closed-form design technique for ultra-wideband Doherty power amplifiers," *IEEE Trans. Microw. Theory Techn.*, vol. 62, no. 12, pp. 3414–3424, Dec. 2014.

[79] J. He, J. Qureshi, W. Sneijers, D. Calvillo-Cortes, and L. Vreede, "A wideband 700-W push-pull Doherty amplifier," in *Proc. IEEE Int. Microw. Symp.*, 2015, pp. 1–4.

[80] A. Barakat, M. Thian, V. Fusco, S. Bulja, and L. Guan, "Toward a more generalized Doherty power amplifier design for broadband operation," *IEEE Trans. Microw. Theory Techn.*, vol. 65, no. 3, pp. 846–859, Mar. 2017.

[81] J. Long, "Monolithic transformers for silicon RF IC design," *IEEE J. Solid-State Circuits*, vol. 35, no. 9, pp. 1368–1382, Sept. 2000.

[82] J. Kim, J. Cha, I. Kim, and B. Kim, "Optimum operation of asymmetrical-cells-based linear Doherty power Amplifiers-uneven power drive and power matching," *IEEE Trans. Microw. Theory Techn.*, vol. 53, no. 5, pp. 1802–1809, May 2005.

[83] D. Kang, J. Choi, D. Kim, and B. Kim, "Design of Doherty power amplifiers for handset applications," *IEEE Trans. Microw. Theory Techn.*, vol. 58, no. 8, pp. 2134–2142, Aug. 2010.

[84] M. Nick and A. Mortazawi, "Adaptive input-power distribution in Doherty power amplifiers for linearity and efficiency enhancement," *IEEE Trans. Microw. Theory Techn.*, vol. 58, no. 11, pp. 2764–2771, Nov. 2010.

[85] W. Chen, S. Zhang, Y. Liu, Y. Liu, and F. Ghannouchi, "A concurrent dual-band uneven Doherty power amplifier with frequency-dependent input power division," *IEEE Trans. Circuits Syst. I: Reg. Papers*, vol. 61, no. 2, pp. 552–561, Feb. 2014.

[86] J. Park, S. Kousai, and H. Wang, "A fully differential ultra-compact broadband transformer based quadrature generation scheme," in *Proc. IEEE Custom Integrated Circuits Conf.*, pp. 1–4.

[87] R. Darraji, F. Ghannouchi, and M. Helaoui, "Mitigation of bandwidth limitation in wireless Doherty amplifiers with substantial bandwidth enhancement using digital techniques," *IEEE Trans. Microw. Theory Techn.*, vol. 60, no. 9, pp. 2875–2885, Sept. 2012.

[88] 89600 VSA Software User Manuals, www.keysight.com.